\documentclass{article}
\usepackage{spconf,amsmath,graphicx,hyperref}
\usepackage{booktabs} % For better tables
\usepackage{amsfonts} % For math symbols like the R in real numbers

\usepackage[final]{changes}

\title{DECAF: Dynamic Envelope Context-Aware Fusion for Speech-Envelope Reconstruction from EEG}

\name{Karan Thakkar \qquad Mounya Elhilali \thanks{This work was supported by NSF 2444353-01 and ONR N00014-23-1-2086.}}
\address{Laboratory for Computational Audio Perception, Johns Hopkins University, USA}
\begin{document}
%\ninept % Uncomment this to use 9-point font if needed for page limits
%
\maketitle
\begin{abstract}
%Decoding a listener's attentional focus from scalp neural recordings is central to applications like neuro-steered hearing aids, but current methods face challenges with fidelity and noise. Prevailing approaches treat this as a static regression problem, decoding each EEG window in isolation and ignoring the rich temporal structure inherent in continuous speech. 

Reconstructing the speech audio envelope from scalp neural recordings (EEG) is a central task for decoding a listener's attentional focus in applications like neuro-steered hearing aids. Current methods for this reconstruction, however, face challenges with fidelity and noise. Prevailing approaches treat it as a static regression problem, processing each EEG window in isolation and ignoring the rich temporal structure inherent in continuous speech. This study introduces a new, dynamic framework for envelope reconstruction that leverages this structure as a predictive temporal prior. We propose a state-space fusion model that combines direct neural estimates from EEG with predictions from recent speech context, using a learned gating mechanism to adaptively balance these cues. To validate this approach, we evaluate our model on the ICASSP 2023 Stimulus Reconstruction benchmark demonstrating significant improvements over static, EEG-only baselines. Our analyses reveal a powerful synergy between the neural and temporal information streams. Ultimately, this work reframes envelope reconstruction not as a simple mapping, but as a dynamic state-estimation problem, opening a new direction for developing more accurate and coherent neural decoding systems.

%Decoding a listener’s attentional focus from scalp neural recordings is central to applications such as neuro-steered hearing aids and brain–computer interfaces, yet remains difficult due to low signal-to-noise ratios and limited fidelity in real-world scenarios. The current study tests whether temporal priors can mitigate these limitations by stabilizing decoding performance over time. We introduce a state-space fusion model that combines neural estimates from EEG decoders with temporal priors predicted from recent speech context, , using a learned gating mechanism that adaptively balances these cues on a time-by-time basis. To isolate the benefits of contextual integration, we evaluate this setup on the ICASSP 2023 Stimulus Reconstruction benchmark using ground-truth priors and show consistent improvements over EEG-only and context-only baselines. Analyses reveal complementary benefits under neural noise and spectral decomposition. Our findings reframe envelope decoding as a dynamic fusion problem, highlighting the potential of contextual priors in robust neural decoding.
\end{abstract}

%Reconstructing attended speech envelopes from neural scalp recordings such as EEG is challenged by low signal-to-noise ratios and limited decoding fidelity in real-world scenarios. Inspired by state-space models, the architecture combines an EEG decoder with a context predictor and adaptively fuses their outputs. 

%
\vspace*{-0.3cm}
\section{Introduction}
\label{sec:intro}

\begin{figure}
    \centering
    \includegraphics[width=1\linewidth]{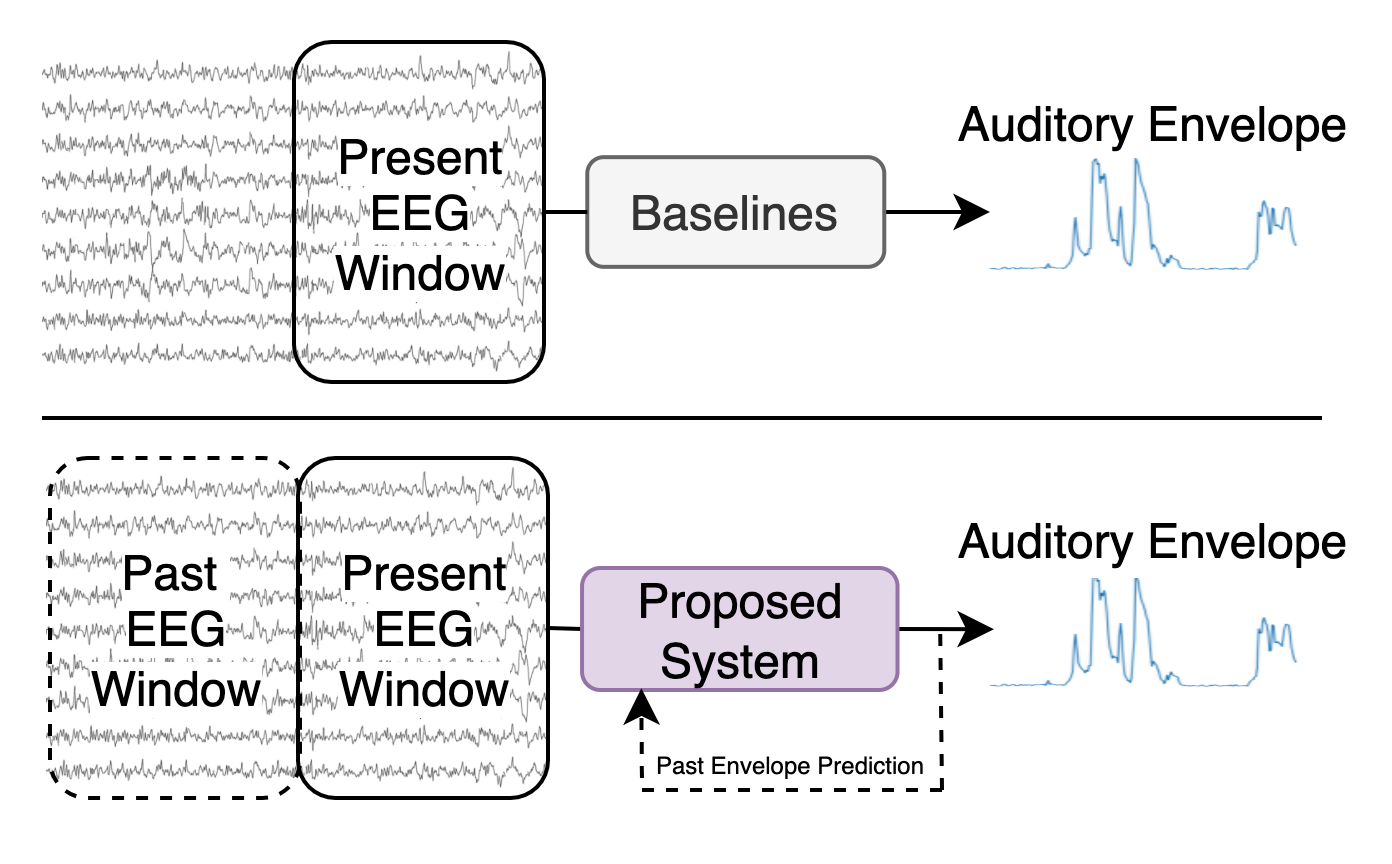}
    \caption{Illustration of the  shift from static to dynamic decoding. (Top) Static baselines perform stateless reconstruction using only an isolated 'Present Window' of EEG. (Bottom) Our dynamic model, DECAF, is state-aware, creating a temporal prior from past context and fusing it with present EEG information. The dashed loop indicates the model's fully recursive operation, using its own past predictions.}
    \label{fig:baseline_vs_proposed}
    \vspace*{-0.5cm}
\end{figure}

Electroencephalography (EEG)-based auditory attention decoding (AAD) aims to determine which speaker a listener is attending to in multi-speaker
environments. A common intermediate task in AAD is reconstructing the envelope of the attended audio stream from EEG signals \cite{O'Sullivan2015Decoding}, which enables downstream correlation-based decoding pipelines. \added{Accurate envelope reconstruction is critical for downstream AAD and neuro-steered hearing aids, where the reconstructed envelope is correlated with audio streams to identify the attended speaker \cite{O'Sullivan2015Decoding, monesi2024auditory}.} Recent deep learning models, such as those based on CNNs \cite{Lawhern2018EEGNet}, LSTMs \cite{Zhang2022}, and Transformers \cite{Xu2022DecodingTransformer}, have significantly improved reconstruction performance \cite{monesi2024auditory}. However, these approaches are often limited by treating envelope reconstruction as a direct mapping problem, where each window of EEG is processed without an explicit, predictive model of the speech envelope's own temporal structure. As illustrated in Fig. \ref{fig:baseline_vs_proposed} (top), each window of EEG is processed in isolation, ignoring the powerful temporal dependencies inherent in a continuous signal like speech. This limitation motivates a critical question: what if we could enhance decoding by creating models that are aware of this temporal context?

The key to creating such a context-aware model lies in leveraging the strong temporal structure of the speech signal itself, which is often available to a BCI system (e.g., via a microphone capturing the audio mixture or recursively from past predictions derived from EEG). This inherent predictability suggests that the task of envelope reconstruction is not static, but fundamentally dynamic. The envelope at any given moment is not an independent event but is highly conditioned by what came immediately before \cite{goupell2019effect, gourevitch2008temporal, o2002linear}. This insight motivates a shift in our approach: from stateless regression to state-aware estimation. \emph{We hypothesize that explicitly modeling this temporal dependency can create a more accurate and coherent decoding system.} The key to this approach is the explicit use of a temporal prior—a prediction of the current envelope generated from its recent past.

\added{To implement this dynamic paradigm, we introduce a state-space fusion model that combines EEG-derived estimates of the current envelope with an autoregressively generated temporal prior from past predictions, inspired by classical signal processing frameworks such as Kalman filtering \cite{kalman1960}. A learned gating mechanism adaptively balances neural evidence and temporal context within a fully causal architecture that relies only on the history of the EEG signal, making it suitable for real-world BCI applications. Our contributions are threefold: (1) we introduce DECAF (Dynamic Envelope Context-Aware Fusion), a novel deep learning architecture that reframes auditory envelope decoding from a static regression problem into a dynamic state-estimation task by integrating direct neural evidence with a predictive temporal prior; (2) the proposed model achieves state-of-the-art performance on Task 2 of the IEEE ICASSP 2023 EEG Decoding public benchmark \cite{monesi2024auditory}, outperforming prior fully causal approaches; and (3) ablation studies and spectral analyses reveal the model’s synergistic fusion of low-frequency neural information with high-frequency temporal structure, leading to improved robustness and decoding accuracy.}

\vspace*{-0.3cm}
\section{Related Works}
\label{sec:related}
The foundational principle behind AAD is that low-frequency cortical responses entrain to the temporal envelope of the attended speech stream \cite{Lalor2010, Ding2012}. This enables reconstructing the attended envelope from EEG and identifying the listener's focus \cite{O'Sullivan2015Decoding}. While early work focused on linear Temporal Response Function (TRF) models \cite{Crosse2016}, recent deep learning architectures such as CNNs \cite{Lawhern2018EEGNet}, LSTMs \cite{Zhang2022}, and Transformers \cite{Xu2022DecodingTransformer, Piao2023HappyQuokka} have become the standard, while newer paradigms such as self-supervised learning are also being explored to learn robust deep representations \cite{thakkar2023SS, }. These models have significantly improved performance on benchmark challenges by capturing complex spatio-temporal EEG patterns \cite{monesi2024auditory}. However, even sequence-aware models like LSTMs and Transformers are typically trained to perform a direct mapping from a window of EEG features to a corresponding speech envelope. They do not, as a rule, explicitly model the autoregressive structure of the speech envelope itself to form a predictive prior.
%The foundational principle behind AAD is that low-frequency cortical responses entrain to the temporal envelope of the attended speech stream \cite{Lalor2010, Ding2012}. This enables reconstructing the attended envelope from EEG and identifying the listener's focus \cite{O'Sullivan2015Decoding}. While early work focused on linear Temporal Response Function (TRF) models \cite{Crosse2016}, recent deep learning architectures such as CNNs \cite{Lawhern2018EEGNet}, LSTMs \cite{Zhang2022}, and Transformers \cite{Xu2022DecodingTransformer, Piao2023HappyQuokka} have become the standard. These models have significantly improved performance on benchmark challenges by capturing complex spatio-temporal EEG patterns \cite{monesi2024auditory}. However, even sequence-aware models like LSTMs and Transformers are typically trained to perform a direct mapping from a window of EEG features to a corresponding speech envelope. They do not, as a rule, explicitly model the autoregressive structure of the speech envelope itself to form a predictive prior.

This direct-mapping approach, without an explicit predictive model of the target signal, overlooks the inherent predictability of speech. The concept of leveraging such priors is a cornerstone of state-aware estimation in other BCI domains. In motor BCI, for instance, Kalman filters use kinematic models to predict hand trajectories, which are then updated by neural evidence \cite{wu2006bayesian, shanechi2016robust}. Similarly, neural speech synthesizers use language models as powerful linguistic priors to ensure the generated output is plausible \cite{anumanchipalli2019speech, ye2025generative}. Inspired by this paradigm, our work applies a similar principle to auditory envelope reconstruction. We propose a state-space fusion model that explicitly generates a temporal prior from the envelope's own recent history and dynamically integrates it with the direct estimate from the EEG, framing reconstruction not as a direct mapping but as an iterative estimation process.

\vspace*{-0.3cm}
\section{Methods}
\label{sec:methods}
\vspace*{-0.3cm}

\subsection{Dataset and Baselines}
\label{ssec:dataset}
\textbf{Dataset:} For this study, we adhere strictly to the protocol and dataset from the ICASSP 2023 Auditory EEG Decoding Challenge (Task 2) \cite{Monesi2023Challenge}. The dataset features 64-channel EEG recordings from 85 subjects listening to narrated stories. We used the officially provided preprocessed data, downsampled to 64 Hz, and the prescribed training, validation, and test splits for all experiments to ensure reproducibility \cite{accou2024sparrkulee}. \added{Evaluation is performed exclusively on the unseen-stimulus test set, which contains novel audio segments from the same subjects used during training, with no overlapping stimulus material.} Following the standard set in the VLAAI paper, we trained our models on 3-second windows of EEG and speech envelope data with an 80\% overlap, and evaluated them on non-overlapping 3-second windows during testing. All models, including our proposed architecture and the baselines, were trained from scratch. The evaluation metric is the Pearson correlation coefficient, which is calculated individually for each subject on unseen test data. The final reported performance is the average of these per-subject scores.

%\subsection{Baselines}
%\label{ssec:baselines}

\textbf{Baselines:} We compare our model against three key baselines spanning from classic methods to the current state-of-the-art. These include a standard multivariate Temporal Response Function model, which uses ridge regression on time-lagged EEG features to provide a simple linear baseline~\cite{Crosse2016}; the VLAAI Network, a deep convolutional network designed for subject-independent speech envelope reconstruction~\cite{Accou2023VLAAI}; and HappyQuokka, the previous state-of-the-art model and winner of the ICASSP 2023 Challenge, which uses a feed-forward Transformer architecture to map EEG signals to speech envelopes~\cite{Piao2023HappyQuokka}.

All decoding models were trained for 10 epochs using an Adam optimizer (batch size of 64) and an early stopping patience of 3. Transformer-based models employed a Noam scheduler~\cite{Piao2023HappyQuokka}, while others used a static learning rate of $1 \times 10^{-3}$. A 500ms delay was applied to the EEG input for all models to account for neural processing time. Further implementation details are available in the public codebase.

\subsection{Proposed Model: DECAF}
\label{ssec:architecture}

\begin{figure}[h!]
    \centering
    \includegraphics[width=\linewidth]{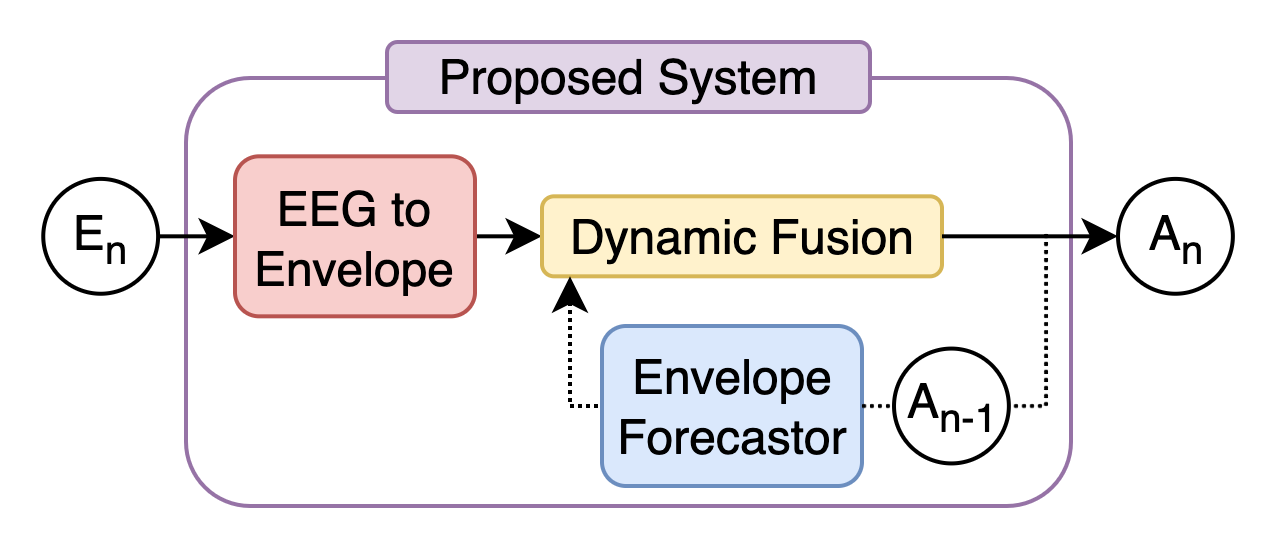}
    \caption{The system generates the current envelope prediction ($A_n$) by fusing a direct neural estimate from the EEG ($E_n$) with a temporal prediction derived from its own past output ($A_{n-1}$).}
    \label{fig:architecture}
    \vspace*{-0.2cm}
\end{figure}

The architecture of Dynamic Envelope Context-Aware Fusion model is designed to implement our dynamic, state-aware decoding paradigm. As illustrated in Fig.~\ref{fig:architecture}, the model reframes envelope reconstruction as a state-space estimation problem, where a prediction from the past is updated by evidence from the present. This is achieved through three core modules that work in concert: an EEG to Envelope decoder, an Envelope Forecaster, and a Dynamic Fusion gate 
\footnote{\added{Training Code: https://github.com/JHU-LCAP/DECAF}}

\begin{table}[t]
\centering
\caption{Speech envelope reconstruction performance on the ICASSP~2023 EEG decoding benchmark (Task~2). Results are reported as mean $\pm$ standard deviation of the Pearson correlation coefficient ($\rho$) across subjects. Improvement is computed relative to the linear baseline.}
\label{tab:performance_summary}
\setlength{\tabcolsep}{4pt} % tighten column spacing
\begin{tabular}{lccc}
\toprule
\textbf{Model} & \textbf{Params} & \textbf{$\rho$ (Mean $\pm$ Std.)} & \textbf{Rel.\ Gain (\%)} \\
\midrule
mTRF (Linear)   & 2.1K  & $0.106 \pm 0.048$ & -- \\
VLAAI          & 6.9M  & $0.153 \pm 0.064$ & +44.3\% \\
HappyQuokka    & 11.1M & $0.162 \pm 0.061$ & +52.8\% \\
\textbf{DECAF} & \textbf{11.4M} & $\mathbf{0.170 \pm 0.061}$ & \textbf{+60.4\%} \\
DECAF-Oracle   & 11.4M & $0.200 \pm 0.048$ & +88.7\% \\
\bottomrule
\end{tabular}
\end{table}

%\begin{figure}[t]
  %\centering
  %\includegraphics[width=\linewidth]{figs/model_comparison_box_plot-1.png}
  %\caption{Baseline vs Proposed. Fusion consistently outperforms EEG-only baselines, with DECAF\_Oracle showing the upper bound performance using the ground truth envelope past as context.}
  %\label{fig:mainresults}
%\end{figure}

\textbf{EEG to Envelope Module:} This module provides a direct, neurally-driven estimate of the speech envelope. It uses the HappyQuokka model's architecture as a powerful feature encoder~\cite{Piao2023HappyQuokka}, which is trained from scratch as part of our end-to-end system. Given a 3-second window of EEG data, represented by the matrix $E_n \in \mathbb{R}^{T \times C}$, where $C=64$ is the number of EEG channels and $T=192$ is the number of time samples ($3\text{s} \times 64\text{ Hz}$), it computes the neural envelope estimate:
\vspace*{-0.2cm}
\begin{equation}
    \hat{A}_{\text{eeg}} = f_{\text{eeg}}(E_n) \in \mathbb{R}^T
\end{equation}
This component can be seen as the ``observation'' in a classic state-space model, representing the evidence provided by the current brain activity.

\textbf{Envelope Forecaster Module:} This module serves as the predictive component of our state-space model, generating a temporal prior. It is a lightweight Gated Recurrent Unit (GRU) \cite{cho2014GRU} encoder that learns the autoregressive structure of speech envelopes. 
\added{The context is first embedded using a 1D convolution (1$\rightarrow$128, kernel size 7)}%
\deleted{It embeds the context using a 1D CNN} followed by a 2-layer 
\added{unidirectional GRU with hidden size 128}%
\deleted{GRU}. 
The GRU's output is then refined by a 
\added{4-head multihead attention mechanism}%
\deleted{4-head attention mechanism}, and a 
\added{feed-forward prediction head maps the final representation to a 3.5~s envelope estimate}%
\deleted{feed-forward head generates the prediction from the final hidden state}. 
Crucially, 
\added{the module operates recursively, taking the model’s own previous output}%
\deleted{its input is the model's own output} from the previous time step, \(A_{n-1} \in \mathbb{R}^{T_c}\) (hereafter, the \emph{context}), making the entire DECAF architecture fully causal 
\added{and recursive}%
\deleted{and recursive:}

\vspace*{-0.1cm}
\begin{equation}
    \hat{A}_{\text{prior}} = \text{Forecaster}(A_{n-1}) \in \mathbb{R}^T
\end{equation}
The output, \(\hat{A}_{\text{prior}}\), acts as a strong temporal prior, the model's belief about the current envelope's shape based on the immediate past.

%\textbf{Dynamic Fusion Module:} The final module adaptively integrates the neural observation (\(\hat{A}_{\text{eeg}}\)) with the temporal prior (\(\hat{A}_{\text{prior}}\)). The goal is to optimally weight the two streams of information to produce the most accurate final estimate. A learnable gating mechanism, implemented as a small MLP, computes a fusion weight \(\alpha \in [0, 1]\) based on the concatenated estimates:

\textbf{Dynamic Fusion Module:} The final module adaptively integrates the neural observation ($\hat{A}_{\text{eeg}}$) with the temporal prior ($\hat{A}_{\text{prior}}$). The goal is to optimally weight the two streams of information to produce the most accurate final estimate. This is achieved with a learnable gating mechanism, implemented as 
\added{a three-layer 1D convolutional network}%
\deleted{3 1D CNN}, which analyzes local temporal patterns in the 
\added{concatenated predictions (2 channels)}%
\deleted{concatenated estimates} to compute a dynamic, time-varying fusion weight $\alpha_t \in [0, 1]$ for each time step. 
\added{Specifically, the gate consists of Conv1d layers with channel dimensions 2$\rightarrow$16$\rightarrow$8$\rightarrow$1, kernel sizes 5, 3, and 1, ReLU activations, and a final sigmoid nonlinearity.}

\begin{equation}
    \alpha = \sigma(\text{Fusion}([\hat{A}_{\text{eeg}}, \hat{A}_{\text{prior}}]))
\end{equation}
The final reconstructed envelope, \(A_n\), is a convex combination of the two estimates, allowing the model to dynamically balance its reliance on direct neural evidence versus its own internal, context-driven predictions:
\begin{equation}
    A_n = \alpha \cdot \hat{A}_{\text{eeg}} + (1 - \alpha) \cdot \hat{A}_{\text{prior}}
\end{equation}

%\textbf{Training Objective:} The entire DECAF architecture is trained end-to-end. The model is optimized using a hybrid loss function that encourages both pointwise accuracy (L1 loss) and global shape matching (Pearson correlation):
%\begin{equation}
%    \mathcal{L} = \lambda_1 \cdot \mathcal{L}_{\text{L1}}(A_n, A_{\text{true}}) - \lambda_2 \cdot \rho(A_n, A_{\text{true}})
%\end{equation}
%Here, \(A_{\text{true}}\) is the ground-truth envelope, \(\rho(\cdot, \cdot)\) is the Pearson correlation, and \(\lambda_1, \lambda_2\) are weighting hyperparameters.

\textbf{Training Objective} 
We train all models with a hybrid loss combining global shape alignment and pointwise accuracy. The Pearson correlation term ($-\rho$) ensures shape similarity, while an L1 term enforces scale correctness. The objective is:
\begin{equation}
    \mathcal{L} = \lambda_1 \, \mathcal{L}_{\text{L1}}(A_n, A_{\text{true}}) - \lambda_2 \, \rho(A_n, A_{\text{true}})
\end{equation}
where \(A_{\text{true}}\) is the ground truth envelope, \(\rho(\cdot,\cdot)\) is Pearson correlation, and \(\lambda_1 = 1,\lambda_2=0.2\) are hyperparameters.

%The EEG encoder maps a 3~s window of 64-channel EEG to a direct neural estimate of the speech envelope using an initial 1D convolution (64$\rightarrow$128, kernel=7) followed by eight pre-layer-normalized FFT blocks (d=128, d$_\text{ff}$=1024, 2 attention heads), and a linear projection to a single output channel with sigmoid activation. In parallel, the envelope forecaster models the autoregressive structure of speech envelopes by processing the model’s own previous output through a lightweight temporal stack consisting of a 1D convolution (1$\rightarrow$128, kernel=7), a 2-layer unidirectional GRU (hidden size 128), and a 4-head multihead attention module, producing a 3.5~s envelope prediction. The final dynamic fusion gate integrates the neural estimate and temporal prior using a three-layer 1D convolutional gating network (2$\rightarrow$16$\rightarrow$8$\rightarrow$1, ReLU + sigmoid) that outputs a time-varying fusion weight, enabling adaptive balancing of EEG evidence and temporal context.

\vspace*{-0.2cm}
\section{Results}
\label{ssec:results}

\begin{figure*}[!t]
    \centering
    \includegraphics[width=1\textwidth]{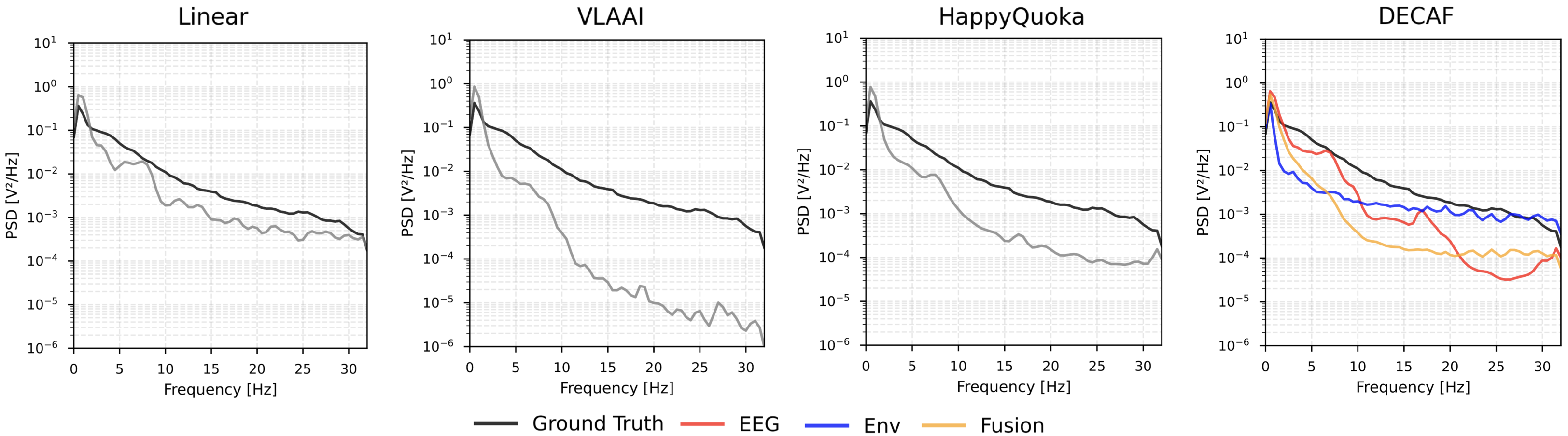}
    \caption{The baseline models (left three panels) effectively capture low-frequency energy but fail to reconstruct higher-frequency details compared to the ground truth (black). The rightmost panel decomposes our proposed model, DECAF. The final Fusion output (blue) synergistically combines the low-frequency accuracy of the EEG branch (red) with the high-frequency information from the Envelope Forecaster (orange), allowing it to track the ground-truth spectrum with significantly higher fidelity.}
  \label{fig:psd}
  \vspace*{-0.5cm}
\end{figure*}

\textbf{DECAF Achieves State-of-the-Art Performance:} As shown in Table~\ref{tab:performance_summary}, our dynamic model, DECAF, achieves a new state-of-the-art in reconstruction accuracy. DECAF ($M = 0.170 \pm 0.062$) significantly outperforms the previous SOTA, HappyQuokka ($M = 0.162 \pm 0.06, $; $p = .000483, d = 0.38$). This performance gain stems from the synergistic fusion of the model's components; an ablation study showed that the EEG branch alone achieved a mean correlation of 0.117, while the Envelope Forecaster branch alone performed near chance ($M = 0.016$), confirming that the final model effectively integrates complementary information from both streams. The model also surpasses other baselines, including VLAAI and a linear mTRF model. Furthermore, we evaluated DECAF\_Oracle, an upper-bound version trained using a ground-truth past envelope as context, which achieved the highest performance (median $\approx 0.20$), demonstrating the full potential of our fusion paradigm.

\textbf{Complementary Spectral Contributions:} 
A Power Spectral Density (PSD) analysis (Figure~\ref{fig:psd}) reveals that while baseline models capture low-frequency neural entrainment ($<$10~Hz), they fail to reconstruct higher-frequency details. DECAF overcomes this through synergistic fusion: its EEG branch captures the low-frequency (1--8~Hz) neural signal, while the Envelope Forecaster's temporal prior preserves the higher-frequency components. By intelligently integrating these complementary streams, the final model restores high-frequency power while retaining low-frequency accuracy, resulting in a spectrally complete envelope that matches the ground truth with high fidelity and explains the model's state-of-the-art performance.

\begin{figure}[t]
  \centering
  \includegraphics[width=\linewidth]{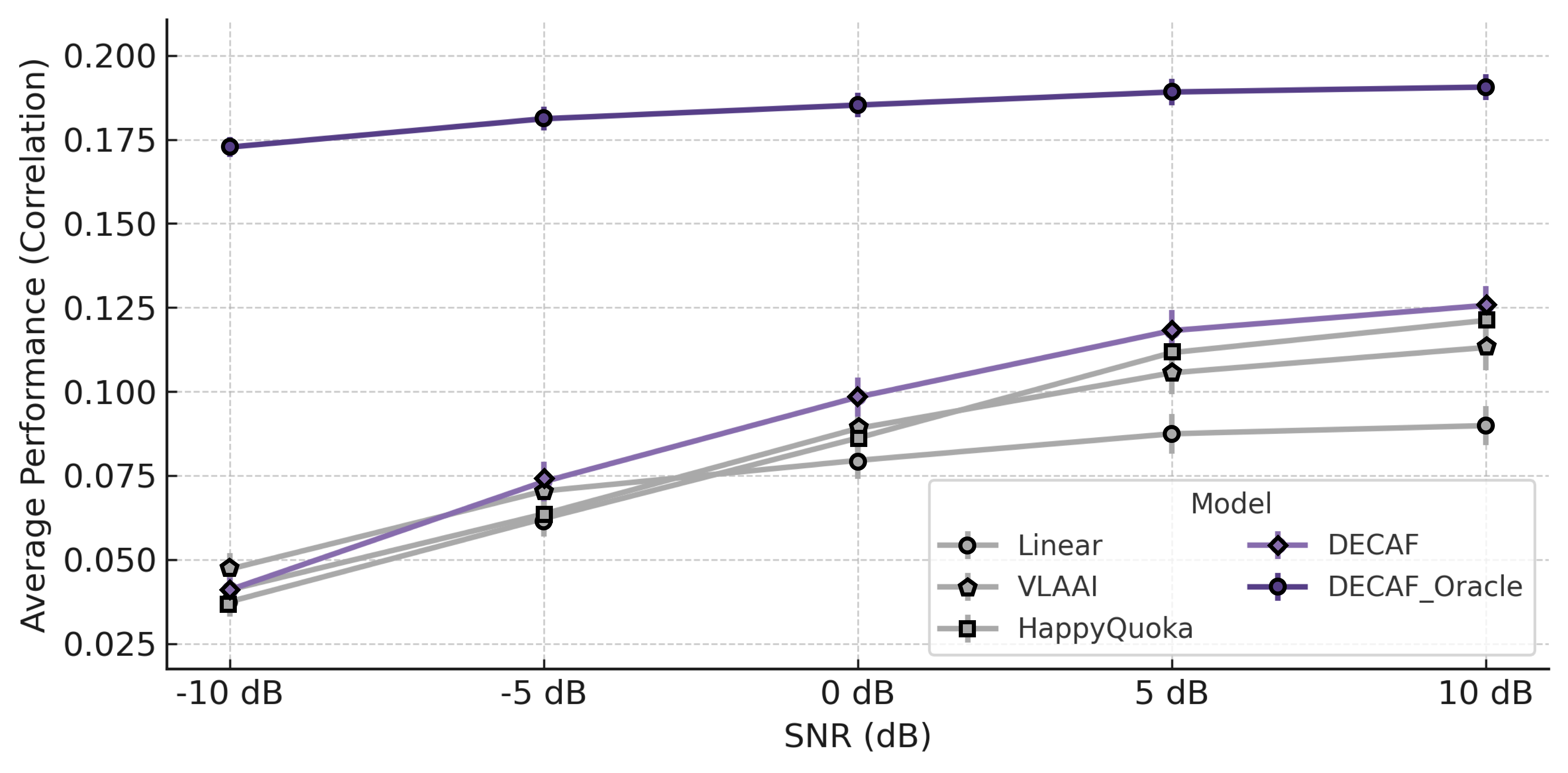}
  \caption{Reconstruction performance across varying EEG noise levels (SNR).}
  \label{fig:robustness}
  \vspace*{-0.6cm}
\end{figure}

%\textbf{Ablation: Effect of White Noise:} To evaluate the model's robustness, a critical feature for practical BCIs, we tested its performance after adding white Gaussian noise to the EEG input at varying Signal-to-Noise Ratios (SNRs), from +10~dB to -10~dB (Figure~\ref{fig:robustness}). While the performance of all baseline models collapses under increasing noise, DECAF exhibits highly adaptive behavior. At high SNR, it leverages the clean EEG to achieve state-of-the-art accuracy. As the EEG becomes corrupted at lower SNRs, DECAF's performance degrades gracefully, eventually stabilizing by relying on its autoregressive temporal prior rather than collapsing entirely like the static models. This provides direct evidence that the dynamic fusion gate learns to intelligently discount unreliable neural input, a key benefit of our state-aware paradigm that makes it highly suitable for real-world applications.

\textbf{Ablation: Effect of White Noise:} To assess robustness under varying input quality, we tested models on EEG data corrupted with additive white Gaussian noise (SNRs: –10 dB to +10 dB). Performance declined with increasing noise, but DECAF showed a clear advantage in how it scales with data quality. At high SNRs (+5, +10 dB), DECAF outperformed baselines by better leveraging clean neural signals. Under extreme noise (–10 dB), this advantage disappeared, with performance aligning to baselines. These results highlight DECAF’s dynamic, state-aware design, which thrives when signal quality is moderate to high. The \texttt{DECAF\_Oracle} provides an upper bound, underscoring the potential of this fusion paradigm for real-world BCIs where signal quality fluctuates.

\vspace*{-0.5cm}
\section{Conclusion and Future Work}
This work was motivated by the hypothesis that making EEG to auditory envelope
decoding models aware of temporal context can yield more accurate and coherent
reconstructions. The results from our DECAF model affirm this, demonstrating how
a recursive temporal prior fosters a synergistic fusion between past context and
present neural evidence for robust envelope reconstruction. \added{The proposed DECAF framework is methodological, modular, and data-agnostic, with a fully causal and recursive design that relies only on past EEG samples and previous model outputs, making it directly applicable to online decoding and closed-loop attention decoding applications.}

\added{This paradigm reframes envelope reconstruction as an iterative inference problem aligned with cortical predictive coding, enabling adaptive, state-aware decoding. Future work will extend this framework to more realistic continuous listening conditions and leverage the learned fusion weights to analyze when decoding relies more strongly on neural evidence versus temporal context. Such analyses can provide insight into the saliency of stimulus segments captured by EEG, revealing how neural and contextual cues contribute differentially across time, and guiding the design of more interpretable and biologically grounded decoding models.}

%This paradigm reframes envelope reconstruction as iterative perceptual inference, mirroring cortical predictive coding, and holds promise for real-time adaptation in neuro-steered interfaces. Future efforts will extend to continuous trials with more explicit brain centric priors, non-linear fusion methods and generative constraints to enhance plausibility, fostering decoding systems that align more closely with the brain's predictive dynamics for naturalistic auditory decoding.

% References should be produced by BibTeX from a .bib file.
% -------------------------------------------------------------------------
\footnotesize
\bibliographystyle{IEEEbib}
\bibliography{refs.bib}

\end{document}